\documentclass[a4paper]{article}

\usepackage{epsf, psfrag, amsmath, amsfonts, amssymb}

\usepackage{graphicx, longtable}

\renewcommand{\thefootnote}{\fnsymbol{footnote}}

\begin{document}
\begin{center}
{\LARGE \bf \sf Optimum ground states of generalized Hubbard models
with next-nearest neighbour interaction\footnote[1]{Work performed
within the research program of the Sonderforschungsbereich 341
(K\"oln-Aachen-J\"ulich)
}}\\[1.5cm]
{\Large \sl Christian Dziurzik, Andreas Schadschneider and}\\[0.1cm]
{\Large \sl Johannes Zittartz}\\[0.8cm]
{\large \sl Institut f\"ur Theoretische Physik, Universit\"at zu K\"oln,
}\\[0.1cm] 
{\large \sl Z\"ulpicher Str. 77, D-50937 K\"oln, Germany}\\[1ex]
{\large \sl (\today)}\\[2cm]  

{\large \sf Abstract}\\[3ex]
\end{center}

{\small We investigate the stability domains of ground states of 
generalized Hubbard models with next-nearest neighbour 
interaction using the optimum groundstate approach. We focus on
the $\eta$-pairing state with momentum $P=0$ and the fully polarized 
ferromagnetic state at half-filling. For these states exact lower bounds 
for the regions of stability are obtained in the form of
inequalities between the interaction parameters.
For the model with only nearest neighbour interaction we show that
the bounds for the stability regions can be improved by considering
larger clusters. Additional next-nearest neighbour interactions
can lead to larger or smaller stability regions depending on the
parameter values. }

\newpage
\renewcommand{\thefootnote}{\arabic{footnote}}

\section{Introduction}
Correlation effects are of great importance in condensed matter physics. 
Superconductivity and ferromagnetism are two important phenomena which can 
arise in an interacting many-body system. Theoretical investigations 
usually begin with choosing a suitable Hamiltonian. In general this 
Hamiltonian is too complex and must be reduced to a reasonable model 
which gives only a simplified description of reality. Such simplifications
make it even more desirable to obtain exact results and compare these 
with experimental data. In addition, they can be used to check the
results from computer simulations and approximative methods.   
\newline
The simplest model of correlated electrons was introduced independently 
by {\sc Hubbard}, {\sc Gutzwiller} and {\sc Kanamori} in 1963 as an 
attempt to describe the effect of correlations for $d$-electrons in 
transition metals \cite{Hub, Gutz, Kan}. This model consists of two terms, 
one describes discrete quantum mechanical motion of electrons ({\em hopping}) 
and the other one the {\em on-site} Coulomb interaction between electrons. 
Nevertheless, the {\em Hubbard model} is one of the most important models in 
theoretical physics and is believed to exhibit various phenomena including 
metal-insulator transition, ferromagnetism, antiferromagnetism and 
superconductivity. In spite of its simplicity only a few exact results 
are known. For instance, {\sc Lieb} and {\sc Wu} solved the one-dimensional 
($D=1$) model by using {\em Bethe-Ansatz}-technique \cite{LW}. The other 
class of exact solutions belongs to the limiting case $D=\infty$, where 
a dynamical mean-field approximation becomes exact \cite{MV, MuHa}. 
However, the situation becomes much more complicated in the lower dimensional 
cases. 
\newline
In recent years a new, non-pertubative method was developed by {\sc
  Brandt} and {\sc Giesekus} \cite{BG}. The main idea is to start with
a well-known ground state and then construct a corresponding
Hamiltonian in the form of a projection operator. This approach
permits to include a large class of interaction parameters. A
generalization of this method was presented by {\sc Strack} and {\sc
  Vollhardt} \cite{SV1, SV2}. {\sc Ovchinnikov} improved some of the
results obtained previously by using a different method \cite{Ovch}
(see also \cite{SdB1}).
His approach is based on {\em Gerschgorin`s theorem} which gives a
lower bound for the ground state energy of the Hamiltonian and thus
complements the usual variational principle which gives upper bounds.
A much simpler and clearer method was used by {\sc de Boer} and {\sc
Schadschneider} \cite{SdB2}. This method is called {\em   Optimum Groundstate 
Approach} and was introduced by {\sc Kl\"umper},
{\sc Schadschneider} and {\sc Zittartz} for spin models \cite{KSZ1}.
The basic idea is to diagonalize a specially chosen local Hamiltonian
and to make all the local states which are needed for the construction
of a given global ground state also local ground states by choosing
the interaction parameters appropriately. This approach leads to some
inequalities between the interaction parameters which represent the
minimal stability region of the investigated ground state. Due to this
restriction only a subspace of the parameter space can be examined.
\newline Using a larger local Hamiltonian enables in a natural way the
inclusion of more interactions which determine the stability
conditions. In general, one finds an extension of the stability domain
of the ground state. Independently, {\sc Szab\'{o}} took this into
account and improved some results obtained previously \cite{Szabo}.
Additionally, he examined the behaviour of the stability domain in the
presence of next-nearest neighbour interaction parameters. For
instance, in the case of $\eta$-pairing state with momentum $P=\pi$ he
verified a shrinking of the stability region for a small ratio between
nearest and next-nearest neighbour hopping. In contrast to his
numerical approach we shall investigate various ground states using
analytical calculations.
    
\section{Method}
A Hamiltonian of a many-body system on an arbitrary lattice but with
homogeneous $\alpha$-nearest neighbour interaction can be split up
into local Hamiltonians, e.g. $H=\sum_{\square}h_{\square}$. The
minimal cluster $\square$ consists of only two nearest ($\alpha = 1$)
neighbour lattice sites $\langle ij\rangle$ and the corresponding
local Hamiltonian is called {\em bond} Hamiltonian. The largest
cluster contains obviously all lattice sites and can be expressed by
$h_{\square}=H$. For small clusters $\square$ the local Hamiltonian
$h_{\square}$ can be diagonalized exactly. This limits the tractable
cluster size.
By adding a trivial constant to the Hamiltonian $H$, which never
changes the physics, one can achieve that the lowest eigenvalue $e_0$
of $h_{\square}$ vanishes, i.e. $e_0=0$. In this case the lowest eigenvalue
$E_0$ of $H$ is either positive ($E_0>0$) or zero ($E_0=0$), because
$h_{\square}$ is a positive-semidefinite operator and the sum of such operators
is also positive semidefinite. In the special case $E_0=0$ a local
condition for finding a ground state $|{\Psi}_0\rangle$ exists:
\begin{equation}
\label{ogs}
H|{\Psi}_0\rangle = 0 \quad \Longleftrightarrow \quad h_{\square}|{\Psi}_0
\rangle = 0 \qquad (\mbox{for all} \quad\square).
\end{equation}
This equivalence can be understood by considering
$\sum_{\square}\langle{\Psi}_0| h_{\square}|{\Psi}_0\rangle =
\langle{\Psi}_0|H|{\Psi}_0\rangle$ $= 0$. Since the $h_{\square}$ are
positive-semidefinite, all $\langle{\Psi}_0|
h_{\square}|{\Psi}_0\rangle$ must vanish, which in turn implies
(\ref{ogs}). In the case $E_0=0$, the global ground state consists
only of ground states of the local Hamiltonian and no excited local
states are involved. A ground state of this type is called {\em
  optimum ground state}. To obtain such ground states for a given
system one must perform two steps. First, the ground states of the
local Hamiltonian must be determined. Then one has to check
whether a global ground state can be formed using only these local 
ground states.

\section{The generalized Hubbard model}
\begin{figure}[]
\centering
\includegraphics[width=0.7\textwidth]{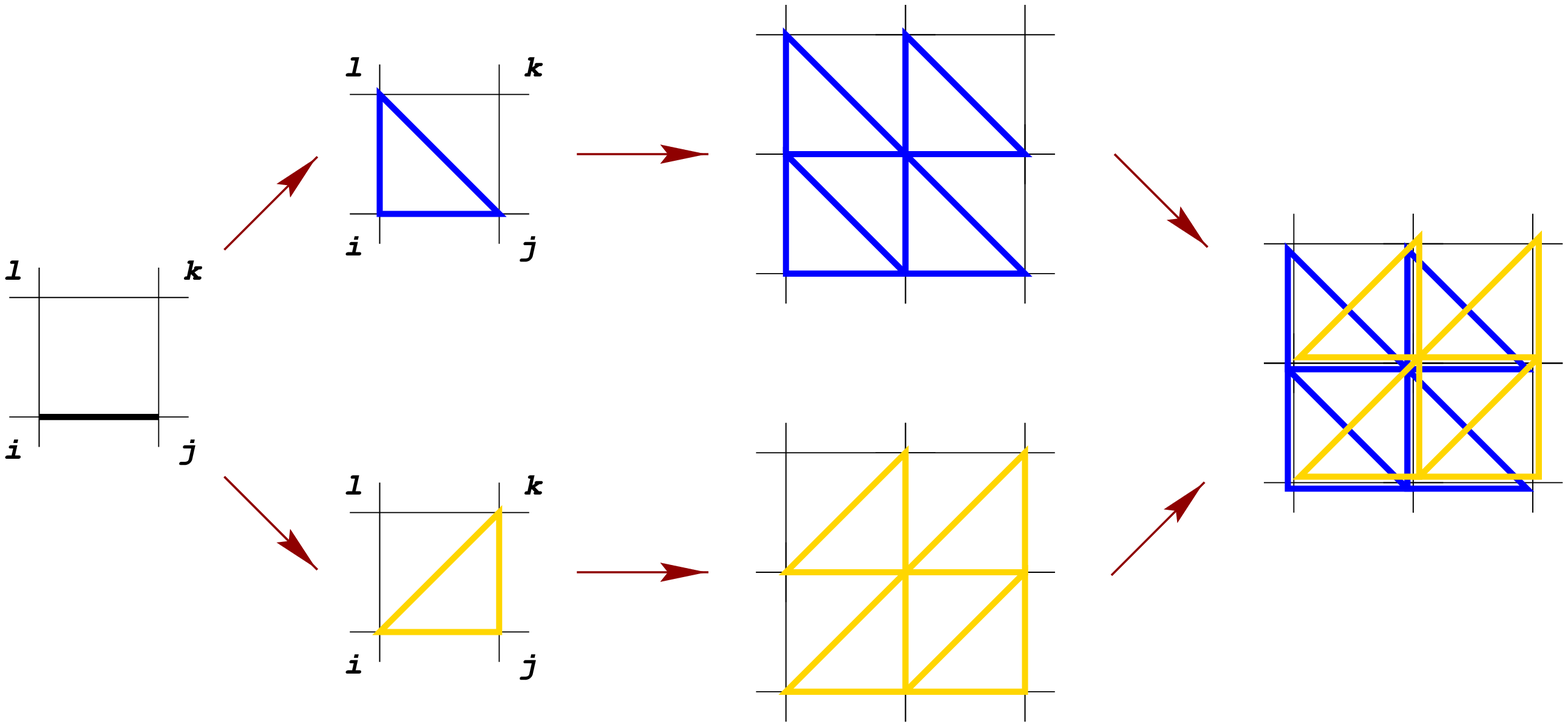}
\caption{\label{tight}{\small {\it Covering of 3-site clusters on a square 
lattice. In order to obtain the full lattice including next-nearest neighbour 
interaction (diagonal bonds) one has to cover the lattice with two different 
{\it 3}-site clusters, namely triangle of type 'lij' (upper part) and 'ijk' 
(lower part).}}}
\end{figure}

The Hamiltonian of the generalized Hubbard-Model on a $D$-dimensional,
hypercubic lattice with $L$ sites and homogeneous $\alpha$-nearest
neighbour interaction can be split up into local Hamiltonians 
$h_{ij}^{(\alpha)}$. Due to homogeneity all local Hamiltonians are 
equal and can be divided into two parts. The first part contains hopping and
interaction terms:
\begin{equation}
    \label{lok-Ham}
    \begin{split}
    h_{ij}^{(\alpha)}= 
    &-{t}_{\alpha}\sum_{\sigma}({\hat c}_{i\sigma}^\dagger {\hat c}_{j\sigma} 
+ {\hat c}_{j\sigma}^\dagger {\hat c}_{i\sigma})\\
    &+{X}_{\alpha}\sum_{\sigma}({\hat c}_{i\sigma}^\dagger {\hat c}_{j\sigma} 
+ {\hat c}_{j\sigma}^\dagger {\hat c}_{i\sigma})({\hat n}_{i, -\sigma} 
+ {\hat n}_{j, -\sigma})\\
    &+{Y}_{\alpha}({\hat c}_{i\uparrow}^\dagger {\hat c}_{i\downarrow}^\dagger
 {\hat c}_{j\downarrow} {\hat c}_{j\uparrow} + {\hat c}_{j\uparrow}^\dagger 
{\hat c}_{j\downarrow}^\dagger {\hat c}_{i\downarrow} {\hat c}_{i\uparrow}) \\ 
    &+\frac{{J}_{\alpha}^{xy}}{2}({\hat S}_{i}^+{\hat S}_{j}^- 
+ {\hat S}_{j}^+{\hat S}_{i}^-)+{J}_{\alpha}^z{\hat S}_{i}^z{\hat S}_{j}^z\\
    &+{V}_{\alpha}({\hat n}_{i}-1)({\hat n}_{j}-1),\\
    \end{split}
\end{equation}
where the pairs $(ij)$ denote $\alpha$-nearest neighbours, for
instance nearest ($\alpha = 1$) and next-nearest ($\alpha = 2$)
neighbours. The fermion operators ${\hat c}_{i\sigma}^\dagger$ and
${\hat c}_{i\sigma}$ create and annihilate electrons
with spin $\sigma\in\{\uparrow, \downarrow\}$ at site $i$ which is
associated with the single tight-binding Wannier orbital. The
corresponding number operators are ${\hat n}_{i\sigma}={\hat
  c}_{i\sigma}^\dagger{\hat c}_{i\sigma}$ and ${\hat n}_i={\hat
  n}_{i\uparrow}+{\hat n}_{i\downarrow}$. The $SU(2)$ spin operators
are given by ${\hat S}_{i}^z=({\hat n}_{i\uparrow}+{\hat
  n}_{i\downarrow})/2$, ${\hat S}_{i}^- = {\hat
  c}_{i\downarrow}^\dagger{\hat c}_{i\uparrow}$ and ${\hat S}_{i}^+ =
{\hat c}_{i\uparrow}^\dagger{\hat c}_{i\downarrow}$. The physical
nature of the various terms is as follows: The first term ($t$) is the
usual hopping of fermions on a lattice. The next two terms,
bond-charge interaction ($X$) and pair-hopping ($Y$), were studied in
relation with superconductivity \cite{Hirsch,Kiv, Pen}. The fourth term is an
anisotropic Heisenberg term with a $XXZ$-type spin interaction given
by the exchange constants $J^{xy}$ and $J^z$. The last term ($V$) is
known as the $\alpha$-nearest neighbour Coulomb interaction. Estimates
for the values of the couplings (for metals) for example can already
be found in Hubbard`s original paper \cite{Hub}. \newline The second
term contains only on-site interactions $O_{ij} = O_{i} + O_{j}$ with
\begin{equation}
O_{i}=\frac{U}{Z}({\hat n}_{i\uparrow}-1/2)({\hat n}_{i\downarrow}-1/2) 
+ \frac{\mu}{Z}{\hat n}_{i}.
\end{equation}
Here $U$ is the on-site Coulomb interaction, $\mu$ the chemical
potential and $Z$ the coordination number of nearest neighbour sites
on the $D$-dimensional hypercubic lattice. \\ 
A local Hamiltonian $h_{\square}$ can be divided into bond
Hamiltonians such that a comparison with the results obtained in
\cite{SdB2} is possible. We restrict our extension to cluster sizes
$N(\square)=\{3, 4\}$ and call the corresponding local Hamiltonians 
{\it 3}- and {\it 4}-site Hamiltonian. In this case only nearest 
($\alpha = 1$) and next-nearest ($\alpha = 2$) neighbour interactions 
exist on the square lattice and therefore:
\begin{equation}
h_{\square}:=\frac{1}{F}\sum_{{\langle ij\rangle}_1\in\square}\left(
h^{(1)}_{ij}+O_{ij}\right) + \sum_{{\langle ij\rangle}_2\in\square}
h^{(2)}_{ij}.
\end{equation}
The factor $F:=2(D-1)$ for $D>1$ is only
necessary in order to compare results of different clusters without
rescaling the coupling constants (due to multiple counts of bonds).
Fig. \ref{tight} shows the
covering of {\it 3}-site clusters on a square lattice.
The entries of a local state are described by $\xi\in\{0,\uparrow,
\downarrow, 2\}$, where '$0$' denotes an empty site,
$\sigma\in\{\uparrow, \downarrow\}$ a site occupied by one electron
with spin $\sigma$ and '$2$' a doubly occupied site. The local state of
cluster size $N(\square )$ is a tensor product
$|\xi_1\xi_2\dots\xi_{N(\square )}\rangle =
|\xi_1\rangle\otimes|\xi_2\rangle\otimes\dots\otimes|\xi_{N(\square
  )}\rangle$. Together one gets $4^{N(\square )}$ local states and a
$4^{N(\square )}\times4^{N(\square )}$-matrix which represents the
local Hamiltonian. Although this matrix might be very large, the
number of zero elements is still a large number. The use of symmetries
makes the problem more tractable. One of the simplest symmetries is
associated with the conservation of the total number of electrons. One
has to consider only subspaces corresponding to a fixed number of
electrons, i.e. one gets a block diagonal matrix. Another useful
condition which we shall frequently impose is $X=t$. It leads to the
preservation of the number of doubly occupied sites (see e.g.\
\cite{SV1,Schatzi}) and thus some of the block matrices split into 
smaller ones. 
Table \ref{tab} summarizes the results for the {\it 3}-site Hamiltonian
with corresponding $64\times 64$-matrix:
\newline
\\
\begin{longtable}{|c|c||c|c|}
\hline
$X\neq t$ & & $X=t$ &\\
\hline
number & block size & number & block size \\
\hline
4 & $1\times 1$   &  4 & $1\times 1$ \\ 
8 & $3\times 3$   & 12 & $3\times 3$ \\
4 & $9\times 9$   &  4 & $6\times 6$ \\  
\hline
\caption{\label{tab}{\it{\small Block sizes and number of blocks for the 
{\it 3}-site Hamiltonian.}}}
\end{longtable}
However, the determination of algebraic eigenvalues of a
characteristic polynomial $p(\lambda)=\det (\mathbb{M}-\lambda
\mathbb{I})$ is limited, i.e. only polynomials up to fourth degree can
be solved in closed form. 
In the case $X=t$ it is possible to find a convenient transformation 
with a corresponding matrix $\mathbb{T}$. After the transformation 
$\mathbb{T}^{-1} \mathbb{M}\mathbb{T} = \overline{\mathbb{M}}$
the four $6\times 6$-matrices decay into blocks of size 3 and all 
eigenvalues can be obtained in closed form. This is the main reason
why we concentrate here on the case $X=t$. 


\section{Results}
We shall restrict our discussion in the following to two 
physically interesting classes of states. 
The first class are the $\eta$-pairing states with momentum $P$ which show
off-diagonal long-range order (ODLRO) and are thus superconducting
\cite{Yang}. The second one is the fully polarized ferromagnetic state
at half-filling. We determine under which circumstances these states
are optimum ground states of the generalized Hubbard model.
We shall mainly consider the square lattice ($D=2$).\\  
The definition of an $\eta$-pairing state with momentum $P$ is given by
the expression
\begin{equation}
      \label{eta-gl}
      |\eta\rangle = \left({\eta}^{\dagger}_P\right)^{{\cal N}}|0\rangle 
\quad \mbox{with} \quad {\eta}^{\dagger}_P=\sum_{j=1}^L e^{iPj}
{\hat c}_{j\downarrow}^{\dagger}{\hat c}_{j\uparrow}^{\dagger},
\end{equation}
where ${\cal N}$ is an integer which is related to the particle number
$N$ through ${\cal N}=N/2$.
Since we would like the $\eta$-pairing state to be the 
ground state of the global Hamiltonian it is informative to determine
the commutator $[H, {\eta}^{\dagger}_P]$. A long, but straightforward 
calculation yields:
\begin{equation}
      \label{kom-gl}
      \small{
      \begin{split}
       [H, {\eta}^{\dagger}_P]=\sum_{\alpha = 1}^2\quad & 2(X_{\alpha}-
t_{\alpha})\sum_{\langle jk\rangle_{\alpha}}\left( e^{iPj}+e^{iPk}\right)
({\hat c}_{j\downarrow}^{\dagger}{\hat c}_{k\uparrow}^{\dagger} 
+ {\hat c}_{k\downarrow}^{\dagger}{\hat c}_{j\uparrow}^{\dagger})\\
       &+2X_{\alpha}\sum_{\langle jk\rangle_{\alpha}}\left(e^{iPj}-e^{iPk}
\right)\left(({\hat n}_{k\uparrow}-{\hat n}_{j\downarrow})
{\hat c}_{j\downarrow}^{\dagger}{\hat c}_{k\uparrow}^{\dagger} 
+ ({\hat n}_{k\downarrow}-{\hat n}_{j\uparrow})
{\hat c}_{k\downarrow}^{\dagger}{\hat c}_{j\uparrow}^{\dagger}\right)\\
       &+\sum_{\langle jk\rangle_{\alpha}}\left(\frac{Y_{\alpha}}{2}e^{iPj}
-V_{\alpha}e^{iPk}\right)({\hat n}_j-1){\hat c}_{k\uparrow}^{\dagger}
{\hat c}_{k\downarrow}^{\dagger}\\
       &+\sum_{\langle jk\rangle_{\alpha}}\left(\frac{Y_{\alpha}}{2}e^{iPk}
-V_{\alpha}e^{iPj}\right)({\hat n}_k-1){\hat c}_{j\uparrow}^{\dagger}
{\hat c}_{j\downarrow}^{\dagger}\\
       &-2\mu{\eta}^{\dagger}_P.
      \end{split}}
\end{equation}
Using (\ref{kom-gl}) one finds the conditions under which the
$\eta-$pairing states (\ref{eta-gl}) are eigenstates of $H$.
For the momenta $P\in\{0, \pi\}$ we have the following constraints
on the interaction constants:
\begin{longtable}{|c|c|}
\hline
$P=0$ & $P=\pi$ \\
\hline
$X_{\alpha}=t_{\alpha} $ & $X_2=t_2$ \\ 
$Y_{\alpha}=2V_{\alpha}$ & $Y_{\alpha}=(-1)^{\alpha}2V_{\alpha}$\\
\hline
\end{longtable}
Note that for momentum $P=\pi$ no conditions relating $t_1$ to $X_1$
exist. In the following we shall only consider the $P=0$ case. 
Other values of the momenta can be treated similarly. An investigation
of the properties of these states can be found e.g. in \cite{Schatzi}. 

\begin{figure}[here]
\centering
\includegraphics[width=1.1\textwidth]{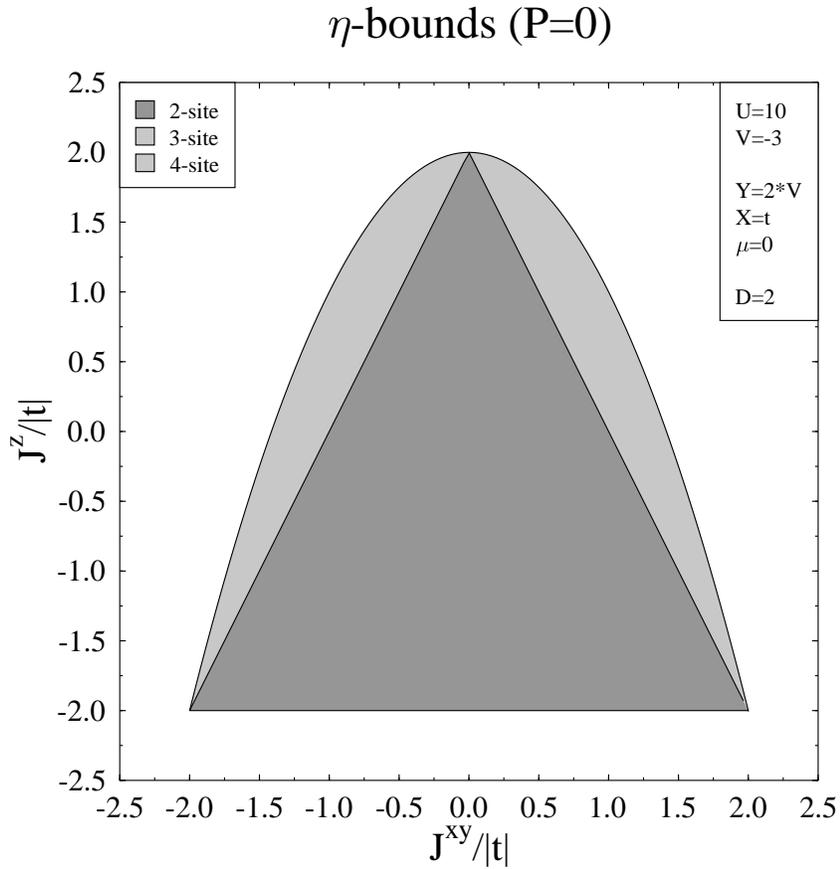}
\caption{\label{eta-bound1}{\it{\small Stability region of the $\eta$-pairing 
state with momentum $P=0$ in the $J^z-J^{xy}$ cut in units of $|t|$ for 
different cluster sizes.}}}
\end{figure}

\begin{figure}[here]
\centering
\includegraphics[width=1.1\textwidth]{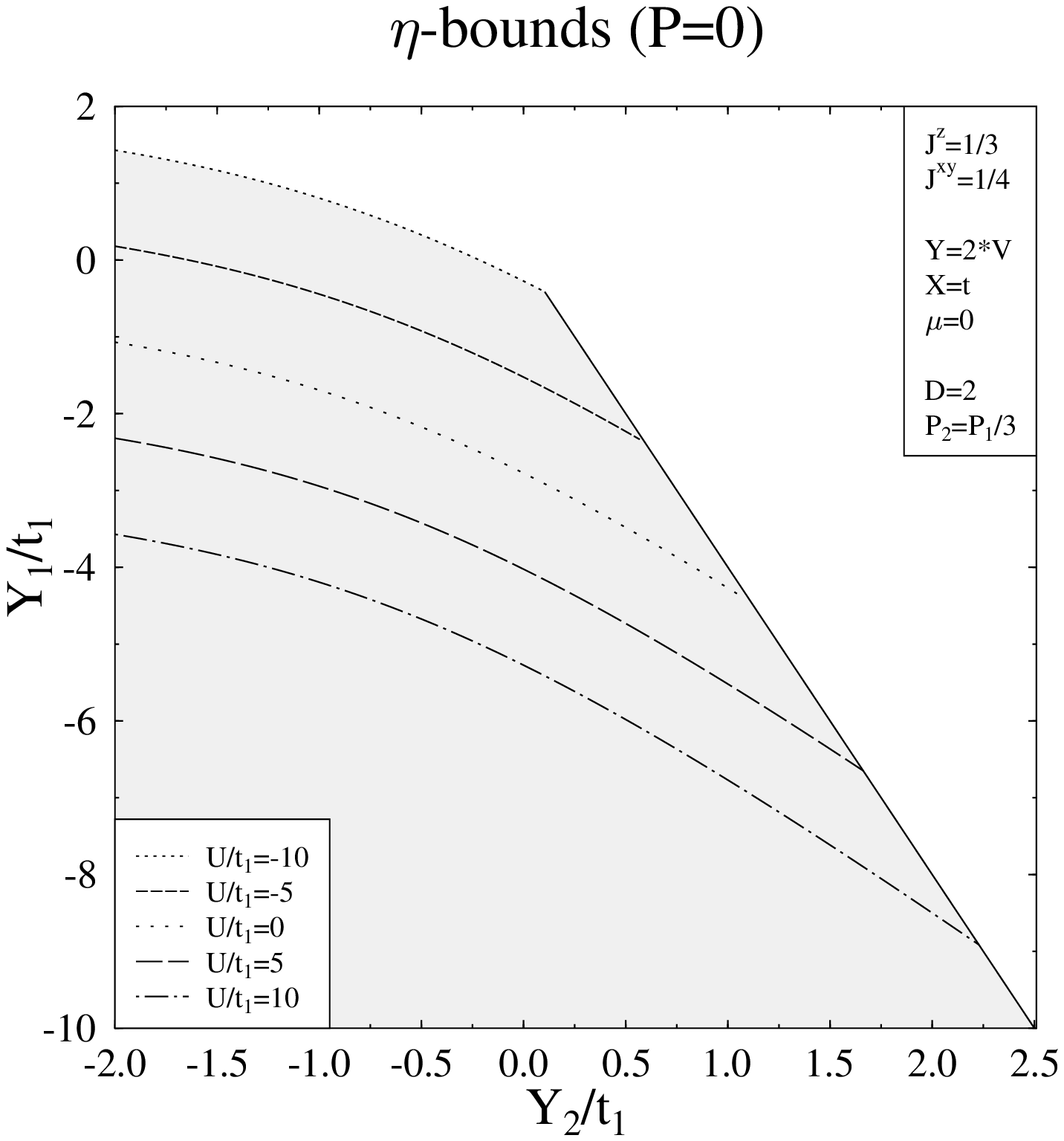}
\caption{\label{eta-bound3}{\it{\small A $Y_1-Y_2$ cut for the bounds of the
$\eta$-pairing state with momentum $P=0$ with non-zero next-nearest 
neighbour interactions. The stability domains are shown for 
different values of $U$.}}}
\end{figure}

In order to make the $\eta$-pairing state (with momentum $P=0$) an 
optimum ground state we first observe that $|\eta\rangle$ can be built 
completely from the local {\it 3}-site states $|000\rangle$, 
$|222\rangle$, $|002\rangle + |200\rangle + |020\rangle$ and 
$|022\rangle + |220\rangle + |202\rangle$ and analogous 
{\it 4}-site states. Without next-nearest neighbour interactions all local
states have the same local energy $e_0=U/(2Z)+V$ if we set $\mu=0$.
Demanding that $e_0$ is the local ground state energy and hence all
other local energies must be larger, one obtains the following
inequalities:
\begin{equation}
      \label{eta-bounds-nN}
      \small{
      \boxed{
      \begin{split}
      & V \leq 0, \qquad
      \frac{U}{Z} \leq\min\left\{\mathfrak{B}_1^{(n)},\, 
\mathfrak{B}_2^{(n)},\, \mathfrak{B}_3^{(n)},\, 
\mathfrak{B}_4^{(n)},\, \ldots\right\}\qquad (n=\it{3},\it{4})\\
      \\
      & \mathfrak{B}_1^{(\it{3},\it{4})} := -2|t|-2V \\
      & \mathfrak{B}_2^{(\it{3},\it{4})} := -V+\frac{J^z}{4} \\ 
      & \mathfrak{B}_3^{(\it{3},\it{4})} := -V+\frac{1}{8}\left(-J^z
-\sqrt{(J^z)^2+8(J^{xy})^2}\right)\\
      & \mathfrak{B}_4^{({\it 3})}       := \frac{1}{3}\left(-5V
-\frac{J^z}{4}-\frac{|J^{xy}|}{2}-\frac{1}{4}\sqrt{\left(4V-J^z-2|J^{xy}|
\right)^2+192t^2}\right)\\
      \end{split}}}
\end{equation}
These inequalities represent the stability regions for the $\eta$-pairing 
state with momentum $P=0$. The selected\footnote{We have listed 
in (\ref{eta-bounds-nN}) only those bounds which are relevant for figures 
\ref{eta-bound1} and \ref{eta-bound2}. A complete list of the 
bounds $\mathfrak{B}_j^{({\it 3})}$ can be found in the appendix. The bounds 
$\mathfrak{B}_j^{({\it 4})}$ can be found in \cite{CeDe}.} bounds 
$\mathfrak{B}_j^{(n)}$ belong to the {\it 3}-site and/or {\it 4}-site case 
and can be distinguished by the upper index $n$. There are seven bounds for 
$n={\it 3}$ and more than $50$ for $n={\it 4}$. The first two bounds were 
also obtained by using the bond-diagonalization \cite{SdB2}. The other bounds 
are new and indicate an improvement of the stability region.\\
It is possible to investigate all cuts of the parameter 
space $(t, U, V,J^z, J^{xy})$, but we shall concentrate only on some of them. 
For all cuts we took realistic parameter values satisfying
$U\geq V\geq t\geq J^z \geq J^{xy}$.\\
In the $J^z-J^{xy}$ cut of the parameter space (Fig. \ref{eta-bound1}) 
the inner triangle corresponds to the stability
region of the $\eta$-pairing state with momentum $P=0$ obtained by
bond-diagonalization. The enlargement corresponds to results achieved
by the {\it 3}-site Hamiltonian (including purely nearest neighbour
interaction). The {\it 4}-site Hamiltonian yields no further improvement 
of the bounds. However, it is not clear that the $\eta$-pairing state is not
a ground state outside of those bounds since larger cluster sizes might yield 
a further enlargement of the stability region.
In contrast to the last figure the following $U-V$ cut (Fig. \ref{eta-bound2})
displays also an enhancement achieved by {\it 4}-site diagonalization.\\
The inclusion of next-nearest neighbour interactions modifies the local 
ground state energy $e_0=U/(2Z)+V_1+V_2$ and hence the constraints concerning 
the interaction parameters:
\begin{equation}
      \label{eta-bounds-nnN}
      \small{
      \boxed{
      \begin{split}
      V_1 &\leq -4V_2, \qquad
      \frac{U}{Z} \leq\min\left\{\mathfrak{B}_1,\, \mathfrak{B}_2,\, 
\mathfrak{B}_3,\, \mathfrak{B}_4, \ldots\right\}\\
      \\
      \mathfrak{B}_1         & := -V_1-V_2+\frac{1}{4}\left(J_1^z
+J_2^z\right)\\
      \mathfrak{B}_2         & := 2\left(-t_2-V_1-V_2
-\sqrt{\left(t_2+V_2\right)^2+t_1^2}\right)\\
      \mathfrak{B}_3         & := \frac{1}{8}\left(-8\left(V_1+V_2\right)
-J_1^z+2J_2^{xy}-\sqrt{\left(J_1^z-2J_2^z+2J_2^{xy}\right)^2
+8(J_1^{xy})^2}\right)\\
      \mathfrak{B}_4         & := \frac{1}{3}\left(-2|t_2|-5\left(V_1
+V_2\right)+\frac{1}{4}\left(J_1^z-3J_2^z\right)+\frac{1}{2}
\left(J_1^{xy}+3J_2^{xy}\right)\right)\\
                             & -\frac{1}{12}\sqrt{\left(-8|t_2|
+4\left(V_1+V_2\right)-J_1^z+3J_2^z+2J_1^{xy}-6J_2^{xy}
\right)^2+192t_1^2}\\
      \end{split}}}
\end{equation}
These bounds belong to the {\it 3}-site diagonalization results since
larger clusters cannot be diagonalized analytically in closed form. 
In order to be close to real systems we take smaller next-nearest neighbour
parameters than corresponding nearest neighbour ones and express this
through ratios $r_P:=P_1/P_2$ with $P_{\alpha}\in\{t_{\alpha},
V_{\alpha}, J_{\alpha}^z, J_{\alpha}^{xy}\}$. The ratios depend on
the material and hence can be very different. Since the $\eta$-pairing
states with momentum $P=0$ consist of electron pairs it is interesting
to consider the $Y_1-Y_2$ cut (Fig. \ref{eta-bound3}). This cut
represents the behaviour of the stability region for different on-site
Coulomb interaction parameters $U$. All other parameter pairs have the
same ratio $r_P=3$. Note that on the square lattice the numbers of 
nearest and next-nearest neighbours are exactly the same. 
One important observation is that the two parameters
$Y_1$ and $Y_2$ stabilize the ground state with increasing Coulomb
repulsion ($U>0$) which try to seperate the electron pairs.
\newline
The fully polarized ferromagnetic state is a simple tensor product of 
local states
\begin{equation}
|\mbox{F}\rangle = \prod_{i=1}^L{\hat c}_{i\uparrow}^{\dagger}|0\rangle,
\end{equation}
where each lattice site is occupied by an electron with spin
$\sigma=\uparrow$ (at half-filling\footnote{Away from half-filling the 
ferromagnetic state is no optimum ground state.}). In contrast to the 
$\eta$-pairing state this state is already an eigenstate of our Hamiltonian 
and hence there are no restrictions concerning the parameters. Nevertheless, 
we concentrate on the special case\footnote{For $X\neq t$ one has to rely 
on numerical methods. Preliminary results show a behaviour similar to the 
case $X=t$.} $X=t$. If we want $|F\rangle$ to become an optimum ground state 
we have to make $|\sigma\sigma\sigma\rangle$ or 
$|\sigma\sigma\sigma\sigma\rangle$ the local ground state. Without 
next-nearest neighbour interaction the corresponding local energy 
$e_0=-U/(2Z)+J^z/4-2\mu/Z$ is a lower bound for the other local energies 
leading to the inequalities:
\begin{equation}
      \label{ferro-gl_nN}
      \small{
      \boxed{
      \begin{split}
      &J^z \leq -|J^{xy}|, \qquad
      \frac{U}{Z} \geq\max\left\{\mathfrak{B}_1^{(n)},\, 
\mathfrak{B}_2^{(n)},\, \mathfrak{B}_3^{(n)},\, 
\ldots\right\}\qquad (n=\it{3},\it{4})\\
      \\
      &\mathfrak{B}_1^{({\it 3}, {\it 4})} := 2|t|+\frac{J^z}{2}+
\frac{2|\mu|}{Z}\\
      &\mathfrak{B}_2^{({\it 3}, {\it 4})} := -V+\frac{J^z}{4}-
\frac{2|\mu|}{Z}\\
      &\mathfrak{B}_3^{(\it{3})} := \frac{1}{2}\left(V+\frac{J^z}{2}+
\frac{|\mu|}{Z}+\sqrt{\left(V+\frac{|\mu|}{2}\right)^2+2Y^2}\right)\\
      &\mathfrak{B}_3^{(\it{4})} := \frac{1}{2}\left(V+\frac{J^z}{2}+
\sqrt{V^2+2Y^2}\right)\\
      \end{split}}}
\end{equation}
The first two boundaries are the same as those of the bond-results
\cite{SdB2}, the last ones are new. Hence an improvement of the
ground state region might be obtained by considering the last two
bounds. But the number of all possible two dimensional cuts is still
very large. 

\begin{figure}[here]
\centering
\includegraphics[width=1.1\textwidth]{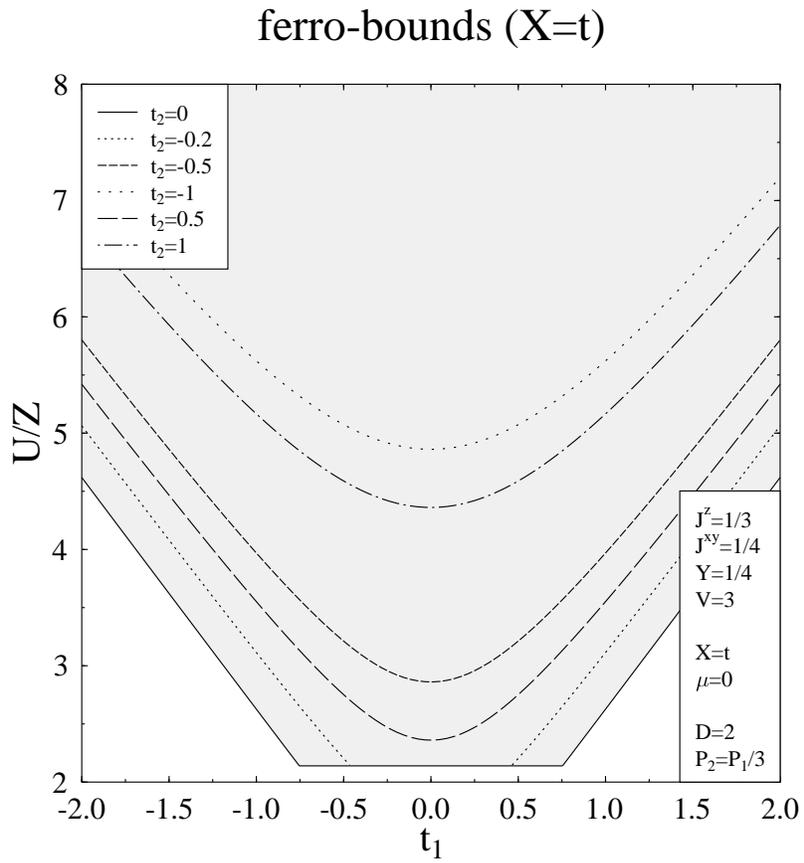}
\caption{\label{ferro-bound1}{\it{\small Bounds for the stability region 
of the ferromagnetic state in the $U-t_1$ plane for different values of
the next-nearest neighbour hopping $t_2$.
}}}
\end{figure}

The bounds can be further improved using the following argument \cite{SdB2}.
With a fixed particle number $N$ a state is a
ground state of $H$ but also a ground state of $H+\mu N$. In this
situation one can regard the bounds as a function of $\mu$ and try to
find the value of $\mu$ which optimizes these bounds. For instance, if
we have inequalities like $a\geq b+\mu$ and $a\geq c-\mu$ then the
best value is $\mu=(c-b)/2$ and thus $a\geq (b+c)/2$. In our case we
get $\mu=0$ and therefore
$\mathfrak{B}_3^{(\it{3})}=\mathfrak{B}_3^{(\it{4})}$. This leads to
the cognition that only the {\it 3}-site diagonalization is necessary,
and the inclusion of four local lattice sites does not improve the
stability region. This result is shown in \cite{CeDe} for various
two dimensional cuts of the parameter space. With next-nearest
neighbour interaction parameters we get the modified local energy
$e_0=-U/(2Z)+J^z_1/4+{J}^z_2/4$ with the corresponding new
inequalities:
\begin{equation}
\small{
\boxed{
      \begin{split}
      J_1^z &\leq -2\left({J}_2^z+{J}_2^{xy}\right), \qquad
      (J_1^z)^2-(J_1^{xy})^2 \leq -2\left({J}_2^z-{J}_2^{xy}\right)J_1^z,\\
      \frac{U}{Z} &\geq\max\left\{\mathfrak{B}_1,\, \mathfrak{B}_2,\, 
\mathfrak{B}_3,\, \mathfrak{B}_4, \ldots\right\}\\
      \\
      \mathfrak{B}_1 &:= -V_1-{V}_2+\frac{1}{4}\left(J_1^z+{J}_2^z\right)\\
      \mathfrak{B}_2 &:= 2{t}_2+\frac{1}{2}\left(J_1^z+{J}_2^z\right)+
\frac{1}{2}\sqrt{\left(4{t}_2+{J}_2^z\right)^2+16t_1^2}\\
      \mathfrak{B}_3 &:= \frac{1}{2}V_1-\frac{1}{2}{Y}_2+\frac{1}{4}
\left(J_1^z+{J}_2^z\right)\\
                     &+\frac{1}{2}\sqrt{\left(V_1+{Y}_2\right)^2+2Y_1^2
-4{V}_2\left(V_1+{Y}_2-{V}_2\right)}\\
      \mathfrak{B}_4 &:=2{t}_2+J_1^z+{J}_2^z-\frac{1}{2}\left(J_1^{xy}
+{J}_2^{xy}\right)\\
                         &+\frac{1}{2}\sqrt{\left(4{t}_2+{J}_2^{xy}
\right)^2+\left(J_1^z-J_1^{xy}\right)\left(J_1^z-J_1^{xy}+2{J}_2^{xy}
+8{t}_2\right)+16t_1^2}\\ 
      \end{split}}}
\end{equation} 
The pure Hubbard model $H=H(t_1, U)$ exhibits no ferromagnetic ground
state at half-filling. Only for some special cases like the Nagaoka
case \cite{Naga} the existence of a fully polarized ferromagnetic
ground state can be proven. An extension to the case of the generalized
Hubbard model can be found in \cite{Kollar}.
The influence of long range hopping
$t_{\alpha}$ on ferromagnetism (at half-filling) was investigated e.g.\ by
{\sc Farka\v{s}ovsk\'{y}} \cite{Far}. The results show
a suppression of ferromagnetism with increasing $\alpha$. In the
$U-t_1$ cut (Fig. \ref{ferro-bound1}) we considered the behaviour of
the stability region for different $t_2$ values. But in contrast to
\cite{Far} all other type of couplings were not turned off.
The inclusion of the next-nearest neighbour hopping shows a reduction
of the stability region for the ferromagnetic ground state in agreement
with the results of \cite{Far}.

\section{Conclusions}
\begin{figure}[here]
\centering
\includegraphics[width=1.1\textwidth]{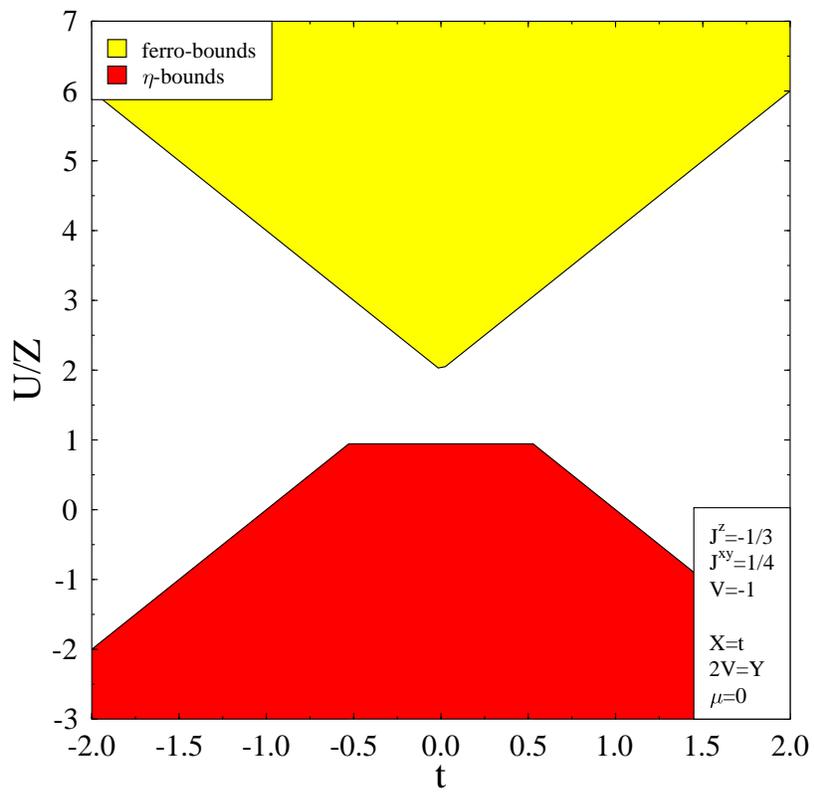}
\caption{\label{vergleich}{\it{\small Phase diagram of the generalized 
Hubbard model with nearest neighbour interactions only.
}}}
\end{figure}

We have presented exact results for stability regions of two physically 
interesting ground states of the generalized Hubbard model with nearest 
and next-nearest neighbour interaction using the optimum ground state 
approach. First we looked at the
$\eta$-pairing state with momentum $P=0$ and then at the fully polarized
ferromagnetic state at half-filling. We have studied the behaviour of the
stability domains of these two states with increasing cluster size. But
due to difficulties that emerge for analytical
diagonalization, we have limited our analytical calculations to two
cluster sizes, i.e. $N(\square)=\{3, 4\}$. These cluster Hamiltonians
were divided into bond Hamiltonians so that a comparison with results
obtained by bond diagonalization \cite{SdB2} was possible. The new
boundaries which exhibit an improvement were illustrated graphically
in some chosen cuts of the parameter space. Without next-nearest
neighbour interactions all cuts show an enlargement of the stability
domains obtained by {\it 3}-site diagonalization. The extension to four 
lattice sites only have led to an amplification in the $U-V$ cut, and all 
other cuts indicated a fast convergence of the stability regions. 
We expect that any further improvement is rapidly decreasing 
with increasing cluster size. Another aim of this work was to determine 
the effects of next-nearest neighbour interactions on the stability domains. 
However, we restricted our investigation to the {\it 3}-site case only. The
illustration of these bounds indicates that the stability conditions are 
strongly dependent on the next-nearest neighbour parameters. For instance, 
in the ferromagnetic case one finds a reduction of the stability 
domain ($U-t_1$ cut) in the presence of next-nearest neighbour hopping 
$t_2$. In the case of the $\eta$-pairing state with momentum $P=0$ we 
considered the
$Y_1-Y_2$ cut for different values of the on-site Coulomb interaction
$U$. With increasing Coulomb repulsion ($U>0$) which tries to separate
the electron pairs we observed a stabilization of the domain because
of large negative $Y_1$ values.
\newline
In summary, we have shown that a cluster of three and four lattice
sites can be treated (with some restrictions) analytically whereby
every two-dimensional cut of the whole parameter space can immediately
be examined and does not require long numerical calculations. 
Although we restricted ourselves mainly to the case $X=t$, where the
local Hamiltonian decays into small blocks which can be diagonalized
in closed form, we like to stress that the optimum ground state approach
can be used to treat the general case $X\neq t$ as well. Here one can
either diagonalize the larger block matrices numerically \cite{Szabo}
or combine the optimum ground state method with the Gerschgorin approach
of \cite{Ovch}. Instead of determining the eigenvalues of the larger
matrices exactly one can obtain lower bounds in closed form by using
Gerschgorin's theorem. Although these bounds will in general not be the
best possible ones they still yield exact stability regions for the
state under consideration.
The results give valuable information about the phase diagram. 
In Fig.\ \ref{vergleich} the stability regions of the two states
investigated here are shown for a generalized Hubbard model with
only nearest neighbour interactions. For the parameter values chosen
one already knows a considerable part of the phase diagram. These
results might serve as a guidance for further computer simulations or
exact diagonalization studies.
Apart from this aspect, the extension enables some interesting new
investigations due to the inclusion of more correlations.\\ 

\section{Acknowledgments}      
We like to thank Jan de Boer and Zsolt Szab\'{o} for helpful discussions.

\section{Appendix}
To obtain an optimum ground state for a given system one has to determine all 
local eigenvalues. For instance, in the {\it 3}-site case (with $\mu = 0$, 
$X_{\alpha}=t_{\alpha}$ and next-nearest neighbour interaction) the 
eigenvalues have the form:
\begin{eqnarray*}
e_1 &:=& \frac{U}{2Z} + V_1 + V_2, \\
e_2 &:=& -\frac{U}{2Z} + \frac{J_1^z}{4} + \frac{J_2^z}{4}, \\
e_3 &:=& \frac{U}{4Z} + t_2 + \frac{V_1}{2}, \\
e_4 &:=& \frac{U}{2Z} - V_2 - Y_2, \\
e_5 &:=& -\frac{U}{4Z} + t_2 + \frac{J_1^z}{8}, \\
e_6 &:=& -\frac{U}{2Z} - \frac{J_2^z}{4} - \frac{J_2^{xy}}{2}, \\
e_{7,8} &:=& -\frac{U}{4Z}\pm{t_2}-\frac{J_1^z}{8}\pm\frac{J_1^{xy}}{4}, \\
e_9 &:=& \frac{U}{2Z} - \frac{V_1}{2} + \frac{Y_2}{2}    
         -\frac{1}{2}\sqrt{(V_1 - 2V_2 + Y_2)^2 + 2Y_1^2}, \\
e_{10} &:=& -\frac{U}{2Z} - \frac{J_1^z}{8} + \frac{J_2^{xy}}{4} 
            -\frac{1}{8}\sqrt{(J_1^z - 2J_2^z + 2J_2^{xy})^2 + 8(J_1^{xy})^2}, \\
e_{11} &:=& \frac{U}{8Z} - \frac{t_2}{2} + \frac{V_1}{4} + \frac{V_2}{2} 
            -\frac{1}{8}\sqrt{\left(-\frac{U}{Z} + 4t_2 -2V_1 + 4V_2 \right)^2 + 32t_1^2}, \\
e_{12} &:=& -\frac{U}{8Z} - \frac{t_2}{2} + \frac{J_1^z}{16} + \frac{J_2^z}{8}             -\frac{1}{16}\sqrt{\left(\frac{2U}{Z} + 8t_2 - J_1^z + 2J_2^z \right)^2 + 128t_1^2}, \\
e_{13,14} &:=& -\frac{U}{8Z}\mp\frac{t_2}{2}-\frac{J_1^z}{16}-\frac{J_2^z}{8}
               \pm\frac{J_1^{xy}}{8}\pm\frac{J_2^{xy}}{4} \\
            && -\frac{1}{16}\sqrt{\left(\pm\frac{2U}{Z}+8t_2\pm J_1^z\mp 2J_2^z-2J_1^{xy}+4J_2^{xy}\right)^2 + 128t_1^2},
\end{eqnarray*}
\begin{eqnarray*}
e_{15,16} &:=& 2\Omega\cos\left(\frac{\theta}{3}\right)-\frac{1}{3}\left(\frac{U}{4Z}\pm 2t_2\mp Y_2-\frac{V_1}{2}-V_2\right), \\
e_{17,18} &:=& 2\Omega\cos\left(\frac{\theta+2\pi}{3}\right)-\frac{1}{3}\left(\frac{U}{4Z}\pm 2t_2\mp Y_2-\frac{V_1}{2}-V_2\right), \\
e_{19,20} &:=& 2\Omega\cos\left(\frac{\theta+4\pi}{3}\right)-\frac{1}{3}\left(\frac{U}{4Z}\pm 2t_2\mp Y_2-\frac{V_1}{2}-V_2\right). 
\end{eqnarray*}
The expressions of the functions $\Omega$ and $\theta$, which include the 
interaction terms, are to large and hence are omitted. \\ 
The bounds are derived from the inequalities $e_0\le e_i$ (for all $i$). For 
the $\eta$-paring state with $P=0$ one has $e_0^{(\eta)} = e_1$. It becomes 
an optimum ground state if the conditions $V \le 0$ and 
$U/Z \le \mbox{min}\{\mathfrak{B}_1, \ldots, \mathfrak{B}_7\}$ are satisfied, 
where
\begin{eqnarray*}
\mathfrak{B}_1^{(\it 3)} &=& -2|t| - 2V, \\
\mathfrak{B}_2^{(\it 3)} &=& -V + \frac{J^z}{4}, \\
\mathfrak{B}_3^{(\it 3)} &=& -V -\frac{J^z}{8}-\frac{1}{8}\sqrt{(J^z)^2 
                             + 8(J^{xy})^2}, \\
\mathfrak{B}_4^{(\it 3)} &=& \frac{1}{3}\left(-5V-\frac{J^z}{4}
                             -\frac{|J^{xy}|}{2}-\frac{1}{4}
                             \sqrt{(4V-J^z-2|J^{xy}|)^2 + 192t^2}\right), \\
\mathfrak{B}_5^{(\it 3)} &=& \frac{1}{6}\left(-8V + J^z\right), \\
\mathfrak{B}_6^{(\it 3)} &=& \frac{1}{6}\left(-8V -J^z + 2|J^{xy}|\right), \\
\mathfrak{B}_7^{(\it 3)} &=& \frac{1}{3}\left(-5V - \frac{J^z}{4}  
                             -\frac{1}{4}\sqrt{(4V+J^z)^2 + 192t^2}\right).
\end{eqnarray*}
Due to the complexity of the last six eigenvalues ($e_{15}$ - $e_{20}$) the 
corresponding bounds do not exist in "closed form", but numerical 
investigations show that they are irrelevant. Including also next-nearest 
neighbour terms one gets: 
\begin{eqnarray*}
\mathfrak{B}_1 &=& -V_1-V_2+\frac{1}{4}(J_1^z+J_2^z), \\
\mathfrak{B}_2 &=& 2\left(-t_2-V_1-V_2-\sqrt{(t_2 + V_2)^2 + t_1^2}\right), \\
\mathfrak{B}_3 &=& \frac{1}{8}\left(-8(V_1+V_2) - J_1^z + 2J_2^{xy} 
                   -\sqrt{(J_1^z-2J_2^z+2J_2^{xy})^2+8(J_1^{xy})^2}\right), \\
\mathfrak{B}_4 &=& -\frac{2t_2}{3} - \frac{5}{3}(V_1+V_2)
                   -\frac{1}{12}(J_1^z + 3J_2^z) + \frac{1}{6}(J_1^{xy} + 3J_2^{xy})\\
                && -\frac{1}{12}\sqrt{(8t_2 - 4(V_1 + V_2) + J_1^z - 3J_2^z - 2J_1^{xy} + 6J_2^{xy})^2 + 192t_1^2}, \\
\mathfrak{B}_5 &=& \frac{2t_2}{3} - \frac{5}{3}(V_1 + V_2) 
                      - \frac{1}{12}(J_1^z + 3J_2^z) - \frac{1}{6}(J_1^{xy} + 3J_2^{xy})\\
                && -\frac{1}{12}\sqrt{(8t_2 + 4(V_1 + V_2) - J_1^z + 3J_2^z - 2J_1^{xy} + 6J_2^{xy})^2 + 192t_1^2}, \\
\mathfrak{B}_6 &=& 4t_2 - 2V_1 - 4V_2, \\
\mathfrak{B}_7 &=& -V_1 - V_2 - \frac{1}{4}(J_2^z + 2J_2^{xy}),
\end{eqnarray*}
\begin{eqnarray*}
\mathfrak{B}_8 &=& -\frac{1}{6}(-8t_2 + 8V_1 + 8V_2 - J_1^z), \\
\mathfrak{B}_9 &=& -\frac{1}{6}(8|t_2|+8V_1+8V_2+J_1^z+2|J_1^{xy}|), \\
\mathfrak{B}_{10} &=& - \frac{2t_2}{3} - \frac{5}{3}(V_1 + V_2) + \frac{J_1^z}{12}                   +\frac{J_2^z}{4}\\                 
                   && -\frac{1}{12}\sqrt{(-8t_2 + 4(V_1 + V_2) + J_1^z - 3J_2^z)^2 + 192t_1^2}.
\end{eqnarray*}
The ferromagnetic state has $e_0^{(F)}=e_2$ and the corresponding bounds can 
be derived in an analogous way.\newline
The number of eigenvalues and bounds in the case $n={\it 4}$ is too large to 
be listed here. They can be found in \cite{CeDe}. 
 
\bibliography{papa}
\bibliographystyle{unsrt}

\newpage  
\begin{figure}[here]
\centering
\includegraphics[width=1.1\textwidth]{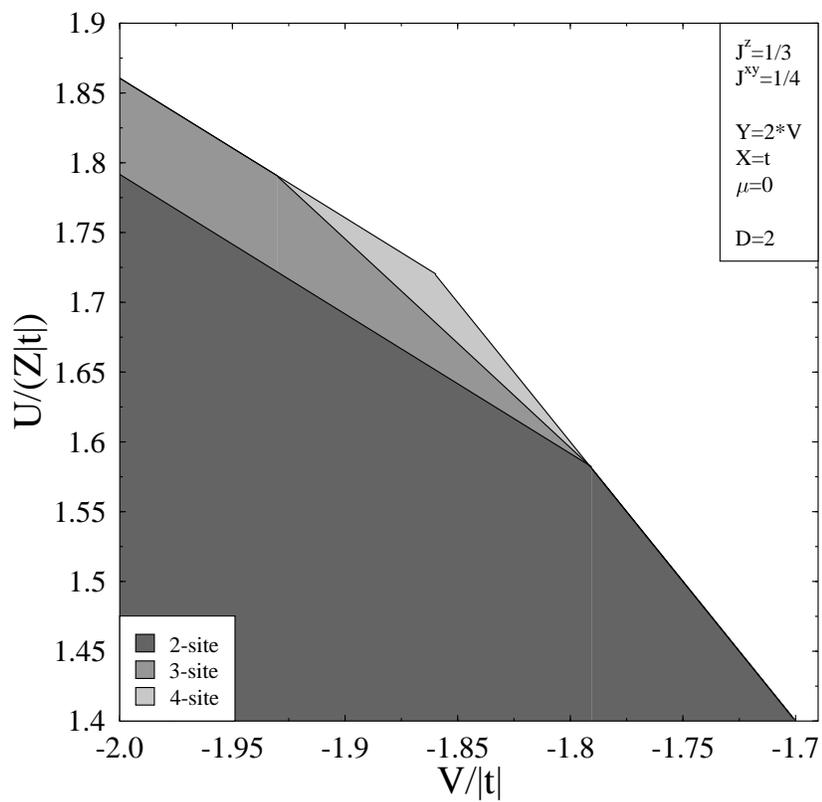}
\caption{\label{eta-bound2}{\it{\small A $U-V$ cut for the $\eta$-pairing 
state with momentum $P=0$ in units of $|t|$.}}}
\end{figure}
\end{document}